# SUBSIDING ROUTING MISBEHAVIOR IN MANET USING "MIRROR MODEL"

Md. Amir Khusru Akhtar[1] and G. Sahoo[2]

[1] Cambridge Institute of Technology, Ranchi, Jharkhand, India
akru2008@gmail.com
[2] Birla Institute of Technology,Mesra, Ranchi, India
gsahoo@bitmesra.ac.in

## ABSTRACT

*Noncooperation or failure to work together is a big challenge that surely degrades the performance and reliability of Mobile Adhoc Networks. In MANETs, nodes have dual responsibilities of forwarding and routing, that's why it needs unison with nodes. To sort out non-cooperation a real life behavior should be implemented, so that misbehavior is nullified.*

*In this paper, we present the "Mirror Model" that strictly enforces cooperation due to its punishment strategy. Node's behavior is watched by its neighbors in PON mode, to update the NPF, NPRF values for a threshold time. After the expiry of the threshold time each node calculates the PFR and broadcasts its neighbors. Similarly all neighbors broadcasted PFR is received and processed by the node to define the 'G' and 'BP' values. The G value is used to isolate selfish nodes from the routing paths and BP denotes the amount of packets to be dropped by an honest node against a selfish node in spite of its misbehavior/packet drops. Cooperation within the neighbors, certainly result in subsiding misbehavior of selfish nodes, therein enhancing cooperation of the whole MANET. This model ensures honesty and reliability in MANET because it does not eliminate a node, but it behaves in the same way as the node behaved. Therefore, it justifies its name, after all mirrors reflects the same.*

*We have implemented the model in "GloMoSim" on top of the DSR protocol, resulting its effectiveness, as compared to the DSR protocol when the network is misconducting for its selfish needs.*

## KEYWORDS

*Mirror, Grade, Bonus Points, Neighbor, Neighborhood*

## ABBREVIATIONS

*Bonus Points (BP), Local Bonus Point (LBP), Grade (G), Number of Packet Forwarded (NPF), Number of Packet Received for Forwarding (NPRF), Packet Forwarding Ratio (PFR), Mobile Adhoc Network (MANET), Misbehavior Gain (MG), Promiscuous Mode Off (POFF), Promiscuous Mode On (PON), Plain DSR (PDSR), Mirror DSR (MDSR)*

## 1. INTRODUCTION

MANETs are the cooperative networks that work without any fixed infrastructure or access point. To impel the correct functioning of the Adhoc network it is really difficult than the wired or infrastructure based network because of the imprudent design. Attacks and dishonest behavior are surely the wall that obstructs the growth and implementation. Models that defend on some threats still face the challenges that we have. Defining a secured, reliable and applicable design that suits mass application is really a big challenge for all. To model the network we have to ensure that the model not only secures the network but also provides reliability and throughput for which the MANET was actually designed.

Our proposed work eliminates misbehaves and enforces cooperation in MANET, due to its stricter punishment strategy called "Mirror Model". Node's behavior is watched by its neighbors using promiscuous listening, to update the NPF/NPRF value for a threshold time. After the expiry of the threshold time each node calculates the PFR value and broadcasts to its neighbors. Similarly all neighbors broadcasted PFR value is received and processed by the node to define the 'G' and 'BP' values. These parameters help to forward a packet by isolating the low grade nodes from the routing paths and thereby to punish selfish neighbor by dropping its packets on the basis of BP value.

In this model each node has to maximize the PFR up to 100% because this ratio is used to define the grade. Therefore, for every packet loss a node will be punished and the punishment cost is in terms of its packet drop by the entire neighbors. The selfish node is punished by an honest node by dropping packets intended for, or originated from, such a node. Routes to be established by bypassing the selfish nodes. A node is punished till the BP is greater than zero and after that the selfish node is automatically added to the network because the BP defines its selfishness. This model does not use any kind of elimination and addition algorithm because a node is punished on the basis of BP.

Nodes become selfish because they have limited resources (such as battery life and bandwidth), that's why the packet dropping behavior would take place. To prevent from the packet dropping attack we have defined BP that denotes the amount of packets to be dropped by an honest node against a selfish node in spite of its misbehavior/packet drops. Thus, a node who wants to save its resources must know that by dropping packets of others, it has to spend more energy to rebroadcast the same packet again and again. Because, its packets are dropped by all its neighbors till the BP value in each node is greater than zero. The punishment cost is substantially more than to behave like an honest node because more energy will be needed to rebroadcast the same packet. So, it knows that selfishness will be harmful, and will be forced to be cooperative.

Use of Mirror Model over DSR [26] protocol to subside the misbehavior can also be used to enhance the DSR protocol by overhearing any communications within its neighborhood. A route reply (RREP) packet can be snooped and a new source route can be added to its route cache. This would minimize the routing overhead incurred due to initiating a route request in further routing. We have deliberately not incorporated this concept in this paper.

This model gives the chances of saving energy to honest nodes by dropping packets of selfish/misbehaved nodes as well as enforces stricter punishment strategy. This model ensures honesty and reliability in MANET, because it does not eliminate a node but it behaves in the same way as the node behaved. Therefore, it shows its name because mirror reflects the same.

The rest of this paper is organized as follows: Section 2 presents background and related work. Section 3 presents the Mirror Model. Section 4 gives the simulation specific assumptions, simulation results and discussion of the results. Section 5 concludes the paper.

## 2. BACKGROUND AND RELATED WORK

Noncooperation is still a challenge for the MANET which limits the expansion and growth. Modification of routing information can be handled by existing secure routing protocols [10-14], but non cooperation is still in its initial stage. To handle non cooperation lots of work [29-41] have been proposed but, they have serious limitation in terms of high overhead.

Since our model is a reputation based model that's why, we are focusing on detection of misbehavior using 'Reputation Based Mechanisms'.

Detection of routing misbehavior was first proposed by Marti et al. [1] named Watchdog and Pathrater. This mechanism was proposed to be used over the DSR [26] routing protocol to mitigate routing misbehavior (including selfish nodes) in ad hoc networks. The watchdog is based on neighbor monitoring for identifying malicious and selfish nodes while Pathrater evaluates the overall reputation of nodes on a path. Pathrater defines route and excludes selfish nodes or misbehaving nodes lying on the paths. In this mechanism selfish nodes are rewarded because there is no punishment for the same. It has another serious drawback that extra battery power consumption because every node has to constantly listen to the medium.

Buchegger et al. [2, 3] introduced the CONFIDANT protocol to observe the behavior of nodes, calculate the reputation of corresponding nodes, and punish the identified selfish nodes. CONFIDANT protocol has four parts: a monitor, a reputation system, a trust manager and a path manager. The Monitor is responsible for recording the behavior information of neighboring nodes. The reputation system is responsible for calculating the reputation of nodes on the basis of direct observation and friends' (indirect) observation. The trust manager is defined to collect warning messages from friends, and the path manager is used to manage routing by excluding selfish nodes. In this protocol, each node monitors its neighborhood behavior and observed misbehavior is reported to the reputation system. If the misbehavior is intolerable then it is reported to the path manager, and then the path manager excludes the nodes from the routing path and calculates new paths. In this regard a warning message will be sent from the trust manager to the friends regarding the misbehaved nodes and after receiving the warning message, the trust manager of a receiver computes the trustworthiness of the message and passes it on to its reputation system if necessary. CONFIDANT has weaknesses in terms of an inconsistent evaluation problem, because in this system each node evaluates different evaluations for the same node and therefore, it is difficult to identify a selfish node. It has another weakness of a location privilege problem because it punishes on the basis of packet dropping not on the basis of how they contributed to the network before. This will drain more battery power for a node situated in the center of the network than the nodes lying on the periphery of the network.

Michiardi et al. [4] proposed reputation measure to know a node's contribution to a network. Reputation is classified into three types: subjective, indirect and functional. Subjective reputation is computed on the basis of node's direct observation, Indirect reputation is computed based on the information provided by other nodes and Functional reputation the subjective and indirect reputation with respect to different functions. It concentrates only routing function and packet forwarding function. After that it takes these reputations to aggregate a collaborative reputation.

Michiardi et al. suggested the CORE [5] protocol to evaluate nodes on the basis of three reputations i.e., collaborative reputation. In this work it maintains a set of Reputation Tables at each node with a watchdog mechanism. The watchdog mechanism is used to observe whether a required function is correctly performed by the requested node by comparing the observed execution of the function with the expected result. The Reputation Table is used to maintain the reputation value of the nodes in the network. Reputation is created/updated along time, based on direct listening by the node itself, or on the basis of information provided by others. This will help a node to judge the selfishness of a service requester and thus decide either to serve or to refuse the request.

Miranda et al. [6] suggested that a node periodically broadcast information's about the status of its neighboring nodes and nodes are allowed to globally declare their refusal to forward messages to certain nodes. This mechanism gives higher communication overhead.

Paul and Westhoff [7] proposed security extensions to the existing DSR protocol to detect attacks in the process of routing. The mechanism depends on neighbor's observations and the routing message's in order to detect the attacker.

"Friends and foes: Preventing selfishness in open mobile ad hoc networks" was proposed by H. Miranda and L. Rodrigues [8]. It employs Friends & Foes to enhance cooperation in MANET. A Friend is able to receive the service of network while Foe's are refused to be served by the nodes of the MANET. The limitation of this mechanism is in terms of memory and message overhead.

"Calculating a node's Reputation in a Mobile Ad Hoc Network" was proposed by W. J. Adams et al. [9]. The proposed Reputation Indexing Window (RIW) method gives stress on current feedback rather than old ones. But, it suffers from serious drawback due to its impractical assumption in which a node is considered as selfish for a long time to build up a good reputation.

In spite of that we have several incentive based mechanisms to enhance cooperation in MANET. Some incentive based mechanisms are as follows

The Packet Purse Model (PPM) / Packet Trade Model (PTM) [15] consist of two models. In the first Packet Purse model the Nuglets are loaded for payment into every data packet before sending it to the intermediate nodes. The intermediate nodes have to obtain more and more Nuglets on the basis of successful forwarding. In the second Packet Trade model the trade packets of intermediate nodes are maintained on the basis of previous nodes. Finally, the destination node assigns credits to all intermediate nodes on the basis of successful transmission of packets. But, the limitation of this model is that it involves secure hardware to control nodes from tampering the amount of Nuglets.

The Adhoc VCG (Vickrey-Clark-Groves) [16] method was proposed by L. Anderegg and S. Eidenbenz. It consists of two phases. The first Route Discovery phase in which the destination nodes compute needed payments for intermediate nodes. After that it intimates this information to the source node or the central bank. In the second Data Transmission phase payment is performed. But, Adhoc VCG mechanism totally depends on the destination node report.

S. Zhong et al., proposed the a Simple, CheatProof, Credit-Based System for Mobile Ad-Hoc Networks [17]. It involves Central Authorized Server (CCS). For a successful forwarding node have to send a receipt to CCS and the CCS assigns credits to nodes according to the receipt. This mechanism suffers from scalability and message overhead.

The "Priority forwarding in Adhoc networks with self-interested parties" was proposed by B. Raghavan and A. C. Snoeren [18]. It has two layered forwarding mechanisms, the Priced Priority forwarding and Free best-effort forwarding. The proposed mechanism suffers from packets forwarding problems.

Protocol independent Fairness Algorithm (PIFA) [19] was proposed by Y. Yoo et al. works for any available routing protocols. It involves a Credit Manager (CM) to maintain the credit databases for nodes, bank stations or sink nodes. But, it also suffers from message processing overhead.

To enforce cooperation several game theory based mechanisms [20-22] and other novel mechanisms [23, 24] have been proposed in the literature, but these mechanisms also suffer from routing and processing overheads.

## 3. PROPOSED WORK

In this section we have presented the "Mirror Model" to enforce cooperation and eliminate misbehavior from MANET. We have implemented the proposed model on the top of DSR protocol [26].

### 3.1. Overview

MANETs may be considered as a society in which nodes agree to cooperate with each other to fulfill the common goal, but non-cooperation is genuine to save itself in terms of their battery power and bandwidth. As we know, cooperation is the basic requirement of MANET that's why we are defining this model that strictly enforces cooperation and eliminates misbehavior. We have defined a grading system to quantify the selfishness of a node. We have used packet forwarding ratio (PFR) as criterion where PFR is the ratio of the number of packets forwarded (NPF) to the total number of packets received for forwarding (NPRF) and it shows a node's contribution to the network.

Every DSR node implements an instance of the 'Mirror Model' and runs in two modes as given below. We have used GlomoSim, a scalable network simulator [28], to simulate our work. At the start of the simulation the MANET must run in protected mode to obtain the Grade and Bonus points to initiate cooperative start.

Normal Mode or POFF mode: in this mode mirror process does not overhear, only regulates its normal activities. It handles/routes packets according to G and BP values on top of the DSR routing protocol. A node has to use Grade and bonus points in the route request, route reply and route error activities. G value is used to forward a packet by isolating the low grade nodes from the routing paths and BP is used to punish selfish neighbor by dropping its packets. A node has to drop the BP amount of packets that results in saving of energy as well as punishing selfish nodes. Modification of NPF and NPRF values are performed in the Protected Mode.

Protected Mode: in this mode mirror process overhears as well as handles/routes packets according to G and BP values on top of the DSR routing protocol. Nodes are supposed to be in a surveillance mode in which all nodes must behave well to obtain a higher grade. Here, we are capturing the parameters for the node such as number of packets received for forwarding, number of packets forwarded at each node in the neighborhood. These values are used to calculate the packet forwarding ratio through which Grade and Bonus Points are defined.

The proposed model has three modules i.e. Neighborhood Behavior Detection Module, Processing, Grading and Bonus Points Allocation Module and Punishment Definition Module. The Mirror Architecture is given in figure 1.

### 3.2. Elementary terminologies

***Definition 1.*** Grade (G): it quantifies the selfishness of a node in its neighborhood.

***Definition 2.*** Bonus Points (BP): it denotes the amount of packets to be dropped by an honest node against a selfish node in spite of it misbehavior/packet drops.

### 3.3. Description of modules

These modules are defined as follows

### 3.3.1. Neighborhood behavior detection Module

This module monitors the neighborhood behavior by promiscuous listing the neighbor traffic. The Mirror process runs on each node for getting information about the neighborhood. It stores the behavior information into a table; Neighborhood Information (NI) Table. This table contains a unique entry for each node of the neighborhood. Nodes have to update the NPRF and NPF values on the basis of number of packets received for forwarding and number of packets forwarded. The schema of the NI table is given below:

NI (IP, NPRF, NPF, G, BP)

where   IP: Internet Protocol Address

NPRF: No of Packet Received for Forwarding

NPF: No of Packets Forwarded

G: Grade

BP: Bonus Points

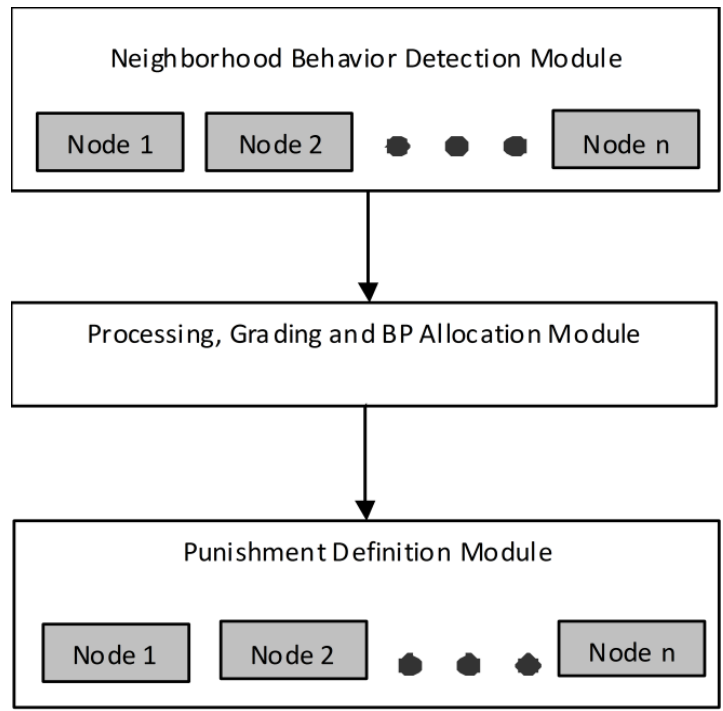

Figure 1. Mirror MODEL

IP field defines the identity of the node in the MANET. Each node monitors its neighborhood behavior by promiscuous listening to increment the NPRF/NPF value of its neighbor. Before incrementing these values a node has to check the BP and if it is greater than zero it shows misbehavior of the corresponding node. The listening node decrements the BP after a packet drop/ allow to drop from the corresponding neighbor and when it is equal to zero then only the increment of the NPRF/ NPF value is performed.

NPRF, NPF, and BP is initialized with zeros and G is initialized with one. The zero value of the given fields indicates the fresh start of the simulation/MANET and we initialize grade with one because this model assumes that an unknown node is honest. The NPRF and NPF values are subsequently updated by overhearing the neighbors on the basis of number of packets received for forwarding and number of packets forwarded.

The Neighborhood Behavior Detection Module working diagram is given in figure 2. In which we have shown 15 nodes arranged in a grid pattern. The nodes, updates its NI table in promiscuous mode, an instance of NI table (for node B) is also shown in figure 2.

| IPA | NPF | NPRF | G | BP |
|-----|-----|------|-----|----|
| A | 9 | 15 | .6 | 4 |
| C | 16 | 20 | .8 | 2 |
| D | 12 | 15 | .8 | 2 |
| E | 13 | 15 | .8 | 2 |
| F | 13 | 15 | .8 | 2 |

Figure 2. Neighborhood Behavior Detection

### 3.3.2. Processing, Grading and BP Allocation Module

After updating the NI table for the threshold time in the promiscuous listening mode, every node has to process the NI table. At first a node calculates the PFR values for its neighbors and broadcast it along with the IP Address to its neighbors. The formula to calculate the PFR is

*PFR = No of Packets Forwarded / No of Packets Received for Forwarding*

Similarly, every node has to broadcast its IP and PFR in its neighborhood, after that these values are collected and filtered by every node of the neighborhood. This information is kept in a Temp Table which is defined as follows.

IT (IP, PFR, G, BP)

where PFR and BP are multivalued attributes, it calculates and store PFR values from NPF and NPRF values. First of all a node writes its own PFR value in the PFR cell and then it appends this field by the received PFR values from its neighbors. An instance of Temp table is shown in Table1.

Table 1.  Temp Table for Node B

| IP | PFR | G | BP |
|----|-----|---|----|
| A | 0.6, 0.8, 0.7 | 0.7 | 3 |
| C | 0.8, 0.7, 0.9 | 0.8 | 2 |
| D | 0.8, 0.7, 0.7, 0.8, 0.6 | 0.7 | 3 |
| E | 0.8, 0.5, 0.7, 0.8, 0.9, 0.9, 0.8, 0.4 | 0.8 | 2 |
| F | 0.8, 0.8, 0.7, 0.7, 0.8 | 0.8 | 2 |

After storing the relevant information in the Temp table nodes have to calculate the Grade and Local Bonus point. The steps are as follows.

Step1: [Calculating Grade for $i^{th}$ node]

The grade is obtained by calculating Mean for the PFR values as

$$G_i = \sum_{k=1}^{n} PFR_k / n$$

where n defines the number of neighbors

Step 2: [Assigning Local Bonus Points for $i^{th}$ node]

$$LBP_i = (MG_i * 10)$$

The LBP is calculated and stored in BP cell locally (in which a node assigns bonus point to its neighbors on the basis of Grade). We are subtracting G from 1 (because $0 \leq G \leq 1$) to find the MG (Misbehavior Gain).

where $MG_i = 1 - G_i$

### 3.3.3. Punishment Definition Module

The punishment is defined on the basis of LBP. After calculating G and LBP for its neighborhood a node has to broadcast the LBP along with the IP address to its neighbors and similarly all broadcasted LBP of its neighbors would be appended in the BP cell of the Temp table. Then the mean value of the LBP is calculated to decide how many packets will be dropped by an honest node against a selfish node in spite of its misbehavior/packet drops. We have calculated the mean value for the LBP cell to make BP value consistent in a neighborhood.

Punishment module is defined as follows

Step 1: [Calculation of BP for $i^{th}$ node]

$$BP_i = \left\lceil \sum_{k=1}^{n} LBP_k / count([LBP]) \right\rceil$$

where count([LBP]) defines the number of values in the respective LBP cell. We have taken ceil of the mean value take BP as integer.

Step 2: [Updating of NI table]

After calculating BP the G and BP column of NI table is initialized/updated by Temp table. The G value helps to forward a packet by isolating the low grade nodes from the routing paths and BP value is used to punish selfish neighbors by dropping BP amount of packets. Thus, the proposed model saves energy of the honest nodes by dropping packets of selfish nodes as well as it provides a stricter punishment strategy to enforce cooperation in MANET.

## 3.4. Adding a new node in the network

In this model inclusion of a new node is very simple, the NPRF, NPF, and BP values are initialized to zero and G is initialized with one. The zero values in the given fields indicate a fresh start of the node in network activities and the value one indicates that our model assumes an unknown node is honest. The NPF and NPRF values are subsequently updated in protected mode by overhearing the neighbors on the basis of the number of packets forwarded or received for forwarding.

## 3.5. Combining Mirror Model with existing Routing Protocols

The proposed model can be easily implemented on top of the existing routing protocols such as DSR [26] and AODV [27]. In DSR protocol, when a node has a packet to send and the destination is unknown, then it has to broadcast a route request (RREQ) packet to its neighbors. Neighboring nodes have to append their own address in the RREQ packet and rebroadcast the RREQ packet to their neighbors. Finally, the RREQ packet is received at the destination and the destination node has to send a route reply (RREP) packet to the source by reverse path.

In order to combine DSR routing protocol with our scheme, packets are handled on the basis of G and BP values. If a RREQ packet arrives from the node whose BP value is greater than zero then it shows misbehavior of the corresponding node. The packet is dropped by the receiving node and therein it decrements the BP by one every time, until it is zero. In the case of overhearing, if a node is watching the traffic of other node then it has to decrement the BP by one after a packet drop from the corresponding neighbor and when it equals to zero then only it increments the NPRF/ NPF value. The G value is used to isolate low grade nodes from the routing paths.

In AODV, whenever a source node needs a route to a destination node, it floods the network with route request RREQ packets. An intermediate node has to reply if it knows a fresh route to the destination, otherwise it propagates the request and nodes update their routing table with a reverse route to the source. When the RREQ reaches the destination, destination replies by sending a RREP towards the source with the reverse route. In the process of route maintenance, upon detecting a link break, a node sends RERR with the active route(s) towards the source(s).

We have used these concepts with our proposed model, by the involvement of G and BP values. A node needs to check the Grade and Bonus Points whenever it gets any RREQ packet. It can drop the BP amount of RREQ packets of the misbehaved nodes. We have applied the same punishment strategy during the route reply and route maintenance. The source can exclude low grade nodes to initiate a new route request. Similarly, if selfish node has to send a RREQ, then it has to spend more energy because its packet has been dropped by its neighbor till BP reaches zero.

## 4. SIMULATION RESULTS

In this section we have discussed the details of the simulation and results.

### 4.1. Details of Simulation Environment

We have used GloMoSim, a scalable network simulator [28] for Simulating Misbehavior in Wireless AdHoc Network. The simulation parameters are listed in Table 2.

Table 2.  Simulation Parameters

| Parameters | Values |
|---|---|
| SIMULATION-TIME | 15M |
| TERRAIN-DIMENSIONS | (1250, 1250) |
| NUMBER-OF-NODES | 121 |
| NODE-PLACEMENT | GRID |
| MOBILITY | RANDOM-WAYPOINT |
| MOBILITY-WP-PAUSE | 30S |
| MOBILITY-WP-MIN-SPEED | 0 |
| MOBILITY-WP-MAX-SPEED | 10 |
| MOBILITY-POSITION-GRANULARITY | 0.5 |
| PROMISCUOUS-MODE | YES |
| ROUTING-PROTOCOL | DSR |

### 4.2. Energy Consumption

In a MANET, nodes have limited resources (such as battery life and bandwidth), that's why energy consumption becomes a major concern. The transmitted power is the strength of the emissions measured in Watts (or milliWatts). A high transmit power will drain the batteries faster, and sensitivity is the measure of the weakest signal that may be reliably heard on the channel by the receiver; the lower value of the signal depends on the receiver hardware performance. Normally values are around -80 dBm, in this model we have used the lowest -90 dBm and obtained good result. In this model to minimize the battery consumption we have reduced the transmitter power because a node has to broadcast or overhear in its neighborhood only. Table 3 shows the parameters used in GlomoSim for Setting up the Transmission Range.

Table 3. Radio Parameters

| Parameters | Values |
|---|---|
| PROPAGATION-LIMIT (dBm) | -111 |
| PROPAGATION-PATHLOSS | Two-Ray |
| RADIO-FREQUENCY (hz) | 2.40E+09 |
| RADIO-TX-POWER (dBm) | 1 |

| RADIO-ANTENNA-GAIN (dBm) | 0 |
|---|---|
| RADIO-RX-SENSITIVITY (dBm) | -91 |
| RADIO-RX-THRESHOLD (dBm) | -81 |
| RADIO RANGE (M) | 125.227 |

## 4.3. Simulation Results

To examine the system we have conducted various tests on a network with selfish nodes. Our underlying protocol for this work is DSR [26], we have implemented as plain DSR (PDSR) i.e., the original DSR and with our enhancement, named Mirror DSR (MDSR).

We have performed our simulation on the following metrics.

### 4.3.1. Packet Delivery Ratio

Figure 3 shows the PDR with respect to percentage of selfish nodes. The PDR decreases as the percentage of selfish nodes increase because more nodes drop packets. Our scheme 'MDSR' shows a higher packet delivery ratio (up to 40%) than the corresponding PDSR. PDR is calculated as

*PDR = Total number of packets received / Total number of packets sent*

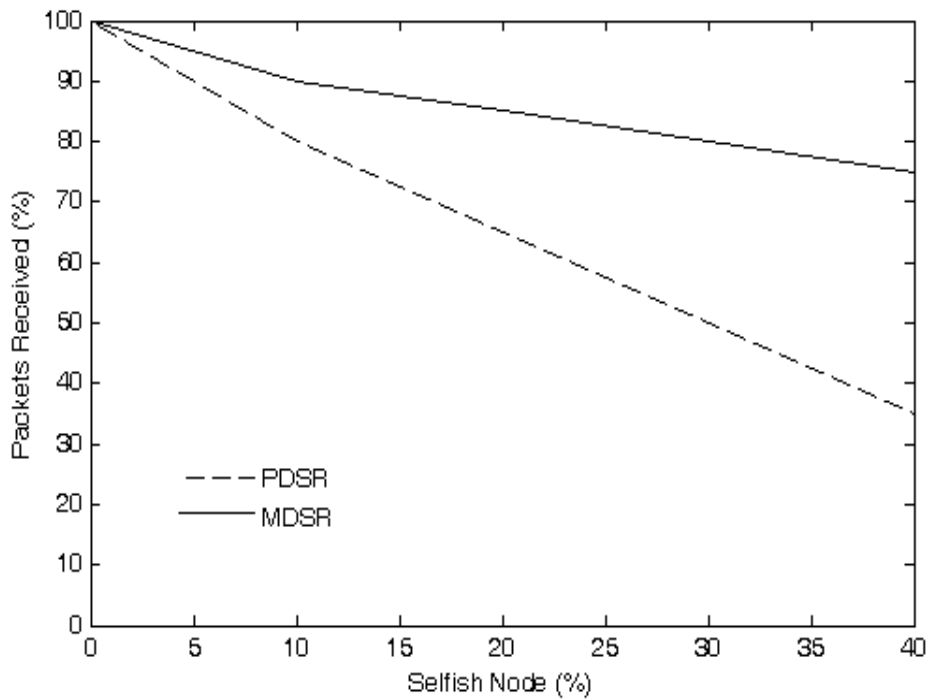

Figure 3. Packet Delivery Ratio

### 4.3.2. Overhead (Number of packets)

Figure 4 shows the overhead in terms of total number of packets with respect to the number of nodes in the network. Laurent Viennot, et al [25] proposed the control traffic overhead for reactive routing protocols, and to guarantee full connectivity in the network one node at least must maintain a route to each node in the network. Implementation of Mirror Model with DSR enhances cooperation with a cost of increase in overhead (minimum of 7.5%).

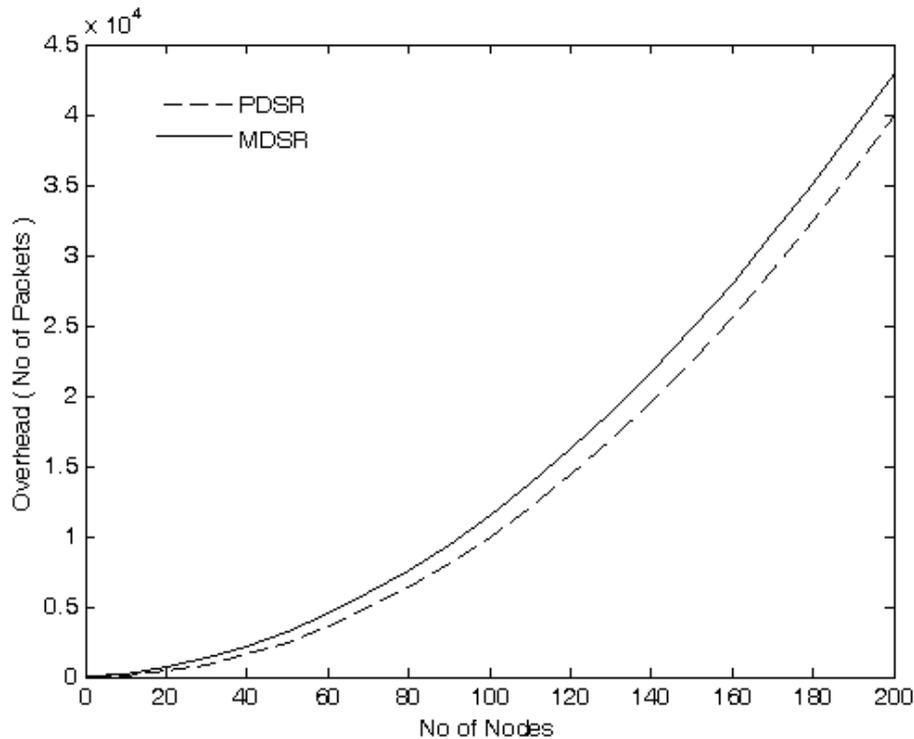

Figure 4. Overhead comparison between a PDSR and MDSR

## 5. CONCLUSIONS

Nodes become selfish because they have limited resources (such as battery life and bandwidth), that's why the packet dropping behavior would take place. To prevent from the packet dropping attack we have used G and BP values. The G value is used to isolate selfish nodes from the routing paths and BP denotes the amount of packets to be dropped by an honest node against a selfish node in spite of its misbehavior/packet drops. Thus, a node who wants to save its resources must know that by dropping packets of others, it has to spend more energy to rebroadcast the same packet again and again. Because, its packets are dropped by all its neighbors till the BP value in each node is greater than zero. The punishment cost is substantially more than to behave like an honest node because more energy will be needed to rebroadcast the same packet. Cooperation within the neighbors, certainly result in subsiding misbehavior of the selfish nodes, therein enhancing cooperation of the whole MANET.

In GlomoSim, we have obtained a higher packet delivery ratio (up to 40%) compared to the plain DSR with a cost of increase in overhead (minimum of 7.5%). In the simulation study we have reduced the transmitter power to save battery power because a node has to broadcast or overhear its neighborhood only. Our model ensures honesty and reliability in MANET because it does not eliminate a node but it behaves in the same way as the node behaved. Therefore, "Mirror Model" justifies its name.

**Authors**

**Md. Amir Khusru Akhtar**

Md. Amir Khusru Akhtar received his M.Tech. in Computer Science. & Engg. from Birla Institute of Technology, Mesra, Ranchi, India in the year 2009 and is pursuing PhD in the area of Mobile Adhoc Network from Birla Institute of Technology, Mesra, Ranchi, India. Currently, he is working as an Assistant Professor in the Department of Computer Science and Engineering, CIT, Tatisilwai, Ranchi, Jharkhand, India. His research interest includes mobile adhoc network, network security, parallel and distributed computing and cloud computing.

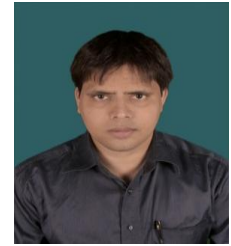

**G. Sahoo**

G. Sahoo received his M.Sc. in Mathematics from Utkal University in the year 1980 and PhD in the Area of Computational Mathematics from Indian Institute of Technology, Kharagpur in the year 1987. He has been associated with Birla Institute of Technology, Mesra, Ranchi, India since 1988, and currently, he is working as a Professor and Head in the Department of Information Technology. His research interest includes theoretical computer science, parallel and distributed computing, cloud computing, evolutionary computing, information security, image processing and pattern recognition.

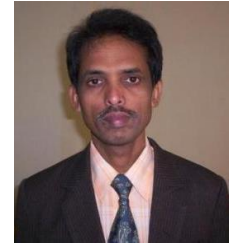